# Galactic-scale macro-engineering: Looking for signs of other intelligent species, as an exercise in hope for our own

*Joseph Voros*[*]

Senior Lecturer in Strategic Foresight
Faculty of Business & Law,
Swinburne University of Technology,
Melbourne, Australia

## Abstract

If we consider Big History as simply 'our' example of the process of cosmic evolution playing out, then we can seek to broaden our view of our possible fate as a species by asking questions about what paths or trajectories other species' own versions of Big History might take or have taken. This paper explores the broad outlines of possible scenarios for the evolution of long-lived intelligent engineering species—scenarios which might have been part of another species' own Big History story, or which may yet lie ahead in our own distant future.

A sufficiently long-lived engineering-oriented species may decide to undertake a program of macro-engineering projects that might eventually lead to a re-engineered galaxy so altered that its artificiality may be detectable from Earth. We consider activities that lead ultimately to a galactic structure consisting of a central inner core surrounded by a more distant ring of stars separated by a relatively sparser 'gap', where star systems and stellar materials may have been removed, 'lifted' or turned into Dyson Spheres. When one looks to the sky, one finds that such galaxies do indeed exist—including the beautiful ringed galaxy known as 'Hoag's Object' (PGC 54559) in the constellation Serpens. This leads us to pose the question: Is Hoag's Object an example of galaxy-scale macro-engineering? And this suggests a program of possible observational activities and theoretical explorations, several of which are presented here, that could be carried out in order to begin to investigate this beguiling question.

---

## Introduction – Big History in context

Big History is a powerful conceptual framework for making sense of the place of humankind in the Universe—a narrative leading from the Big Bang nearly 14 billion years ago to our present information-based technological civilization (e.g., Brown 2008; Christian 2004, 2008; Spier 1996). It synthesises many different knowledge domains and scholarly disciplines and, in the words of the International Big History Association (IBHA): 'seeks to understand the integrated history of the Cosmos, Earth, Life and Humanity, using the best available evidence and scholarly methods' (International Big History Association 2012).

Nonetheless, Big History is ultimately concerned with the history of just *one* planet—ours—among the trillion or so that are now thought to exist in the Milky Way Galaxy, not to mention the billions of trillions that can thereby be inferred to exist in the wider observable universe. Thus, it can be considered a single case in the even larger context of the unfolding of the broad scenario of Cosmic Evolution, as that scenario has played out on this particular planet (e.g., Chaisson 2001, 2007, 2008; Delsemme 1998; Jantsch 1980). It is easy to imagine other planets where life, and perhaps even intelligence, has arisen, as the Cosmic Evolutionary scenario has unfolded there, possibly giving rise to their own unique variant of Big History. A natural sub-set of the study of Cosmic Evolution is the discipline of Astrobiology (e.g., Chyba and Hand 2005; Mix et al. 2006), the study of life in the universe, which also includes its own associated sub-set of SETI, the search for extra-terrestrial intelligence (e.g., Ekers et al. 2002; Harrison 2009; Morrison, Billingham and Wolfe 1979; Sagan and Shklovskii 1966; Shostak 1995; Tarter 2001). So, one can imagine an expanding set of nested fields of study, beginning with Big History—the history of our own small 'pale blue dot' (Sagan 1995)—enfolded by Astrobiology/SETI—the study of how life may arise in the universe, and the search for intelligent forms of it—and encompassed by Cosmic Evolution—the study of how our universe as a whole has changed over the course of deep cosmic time. Whether there is a further enfoldment of our own universe within an even larger 'multiverse' of other universes is a fascinating open question currently receiving some attention among cosmologists.

Our focus here will be on using the step beyond Big History—specifically SETI—as a framework for thinking about some of the broad contours that might characterize the unfolding future for intelligent technological civilizations, including—possibly—our own. Searching for signs of long-lived intelligent extra-terrestrial species—such as those that will be sketched here—could provide one way for us to shift our collective thinking to become much more far-reaching and much, much longer-term—something that increasingly appears to be vitally necessary for the future of our civilization and planet. And if we actually can begin making this collective worldview change—albeit perhaps only minutely at first—then even this small shift of our current mindset could become a rational basis for some measure of hope in our ability to determine both wisely and well what the next stages will be in the long-term Big History view of the potential 'future histories' of our species.

## Approaches to the Search for Extra-Terrestrial Intelligence

The modern search for extra-terrestrial intelligence has a history of just over half a century (Dick 2006). It has mainly involved searching for electromagnetic signals, usually at radio frequencies, although, more recently, it has also been undertaken in the optical spectrum (Shostak 2003). Just after the original proposal by Cocconi and Morrison (1959) to search for radio transmissions, Freeman Dyson (1960) suggested looking not for electromagnetic signals but instead for artificial signs of technology, an idea elaborated in further detail a few years later (Dyson 1966). Thus, some recent SETI researchers (e.g., Bradbury, Ćirković and Dvorsky 2011; Ćirković 2006) consider there to be two main approaches to SETI: the *'orthodox'* approach, based on the detection of electromagnetic signals, whether they were deliberately signalled or are simply unintentional 'leakage' from the civilization; and the *'Dysonian'* approach, based on looking for signs of extra-terrestrial technology, without any presumption of deliberate signalling or attention-seeking at all.

The idea of searching for signs of extra-terrestrial technology or artefacts (e.g., Freitas 1983) is an example of an approach that has more recently been called 'interstellar archaeology' (Carrigan 2010, 2012). Such a form of archaeology is hampered, of course, by the enormous distances to other stars, making the more usual pick-axe and soft-brush approach impractical (to say the least!), so any examples of technologies or artefacts we would be able to discover in this way would probably need to be executed on a stupendous scale. Some researchers have suggested, however, that we might possibly find artefacts in our own solar system (e.g., Kecskes 1998; Papagiannis 1983). Unfortunately such an enticing 'field trip' is currently beyond our technical ability, although there have been some ideas proposed for the exploration and use of near-Earth objects, including asteroids, in this next half-century (e.g., Huntress et al. 2006). Along similar lines, Davies and Wagner (2013) have recently suggested taking a closer look at the Moon for any possible traces of extra-terrestrial technology.

The emerging field of 'macro-engineering' is explicitly concerned with thinking about engineering on large scales, so it may be helpful as a framework for informing our thinking in the search for insights regarding examples of extra-terrestrial technology (Badescu, Cathcart and Schuiling 2006; Cathcart, Badescu and Friedlander 2012). Macro-engineering can be conceived of at a variety of scales, ranging from sub-regional to planetary to stellar to galactic in scope (Badescu et al. 2006). In what follows below we consider the last of these—galaxy-scale macro-engineering—and how we might go about imagining what forms such almost unimaginable feats of engineering might take, in order to think about how we might detect such artificial activities across intergalactic distances. But first let us examine more closely why this type of approach to SETI may be even more relevant today than it has conventionally been considered to be.

**The Drake Equation**

One of the pioneers of SETI, Frank Drake, developed an equation which has since become widely used as a conceptual framework for discussion and debate about the various terms which are included in it (Drake 1961). The Drake Equation can be written as:

$$N = R_* \times f_p \times n_e \times f_l \times f_i \times f_c \times L$$

where $N$ is the number of currently-existing communicating technological civilizations in the Milky Way Galaxy; $R_*$ is the average rate of formation of suitable stars per year in the Galaxy; $f_p$ is the fraction of these stars which have planets; $n_e$ is the average number of planets in each of these star systems with conditions favourable to life; $f_l$ is the fraction of these planets which go on to actually develop life; $f_i$ is the fraction of these inhabited planets which go on to develop intelligent life; $f_c$ is the fraction of planets with intelligent life that develop technological civilisations which are capable of releasing signals into space; and $L$ is the average communicative lifetime of such a civilisation. There are several variants to this equation, and there have been many modifications made to it over the ensuing decades as well (e.g., Bracewell 1979; Ćirković 2004; Hetesi and Regály 2006; Maccone 2010; Walters, Hoover and Kotra 1980).

With regard to the parameter $L$, initially this was often taken to mean the actual lifetime of the civilization. Some early estimates of this parameter tended to be rather gloomy, therefore, given our own case of the unwelcome possibility of hair-trigger nuclear annihilation under which humanity has lived since the middle of the 20$^{th}$ Century CE. It was often thought, based on our own example, that many nascent technological civilizations might therefore destroy themselves not long after achieving the ability to send signals into space. More recently, as suggested by the above, the meaning of $L$ has shifted subtly from the 'lifetime of the civilization' to the 'length of time such civilizations release detectable signals into space', a shift which changes the character of the term rather significantly. This newer meaning for $L$ has again arisen through reasoning from our own example. The Earth's radio 'signature' has changed over the decades from very high-energy analogue broadcasts from ground-based transmitters aimed towards the horizon (where the suburbs and home viewers are) and which have thereby continued on further out into space, to very low-energy narrow digital beams from orbiting satellites that are targeted at particular regions of the Earth's surface, and therefore do not propagate to any great degree beyond the Earth. This shift from high-power analogue broadcasts to low-power digital 'narrowcasts' (so to speak) has meant that over time the Earth is becoming less and less visible due to 'leakage' from our own terrestrial use of radio waves (Drake 2010).

SETI commentators sometimes use the US TV show *I Love Lucy* as the archetypal example of the sort of broadcast material that is expanding outwards in a 'bubble' from the Earth at the speed of light as a 'cultural leakage' signal from our civilization. One might be moved to observe that, if *those* are the signals that form the basis of an assessment by extra-terrestrials as to whether there is intelligent life on Earth, then perhaps it is *no wonder* that extra-terrestrials have not sought to make contact! Whatever one's opinion about the value of the content it carries, however, over time this leaked electromagnetic energy will die down to a faint whisper, due to the changing pattern of electromagnetic radiation use on Earth. Eventually, once the initial several decades' worth of high-energy broadcasting has passed, the longer-term low-power digital signal emanating from Earth will likely become almost impossible to detect by chance above the normal background radio noise of space. Only some navigational beacons and radar are likely to remain detectable after this time.

This gradual disappearance over time of the Earth as a strong radio source has implications for the wider consideration of the value of $L$ in the Drake Equation. It is possible to imagine that other civilizations might also make a similar such transition, so that a civilization might indeed be very long-lived, even while having a value of $L$ remaining relatively short (Drake 2010). This implies a need to re-think some of the conventional approaches to SETI, or at least some of the assumptions upon which orthodox SETI has been based for so long, if not to expand the thinking beyond conventionality altogether. Indeed, Frank Drake himself has commented that (Foreword to Shostak 2009):

> Searching for extra-terrestrial signals is one of the most challenging tasks ever taken on by mankind. … We are challenged to use logic to predict what another civilization, probably much older and more advanced than us, might adopt as a technology we might detect. … To reach an answer, **we have to become futurists**, *reaching far beyond our usual comfortable world of telescope technology to arrive at possible scenarios for the distant future*. This becomes an exercise of intellect reaching far beyond the usual bounds of science theory. (Emphases added.)

This then is the challenge, as posed by one of the founding pioneers of SETI: to imagine how an advanced civilization might develop over the course of perhaps hundreds of thousands or even millions of years, and attempt to conceive of what sort of technology such a civilization might invent or use. We are challenged, in other words, to think on a truly 'cosmological' scale, thinking which will very likely need to include both temporal and spatial dimensions—vast distances and immense timeframes. In order to approach and meet this challenge, we will need some sort of organizing principle for doing so.

**'Dysonian' Thinking**

Fortunately, Dyson explicitly set out the three 'rules' for his 'game' of thinking about extra-terrestrial technology (Dyson 1966). With very little modification, it is possible to use them from the point of view of our current understanding of technology, and with a view to incorporating ideas coming from the above-mentioned field of macro-engineering. Dyson's three rules can be given as follows (Dyson 1966: 643-644):

1. Think of the biggest possible artificial activities, within limits set only by the laws of physics, and look for those;
2. All engineering projects are carried out with technology which the human species of the *current epoch* can understand; and
3. Ignore questions of economic cost;

and where, in Rule 2, the italicized term 'current epoch' replaces Dyson's original use of 'year 1965 AD'. The given modification allows the rule to be applied at any stage of human history, which will thereby yield different answers depending upon the state of our knowledge in any given epoch. In particular, Dyson stressed with respect to Rule 1 that he was not interested in what an 'average' technological civilization might look like, only in what the most conspicuous of perhaps one in a million might look like, as these would be the easiest to detect over great distances; hence the focus on the 'biggest possible artificial activities'.

Dyson went on to outline how it would be possible to disassemble planets, build rigid structures in space, and also revisited some of the ideas in his earlier paper (1960) which suggested that attempts to harvest increasing amounts of stellar radiation from the civilization's home star would lead, in the asymptotic limit, to all visible radiation being intercepted by a vast 'swarm' or 'shell' of orbiting collectors completely enveloping the star. While no longer radiating in the visible spectrum, such an object would nonetheless remain visible in the infra-red, owing to the black-body radiation law, whence Dyson's proposal in the initial paper's title to search for artificial sources of infra-red radiation. There have been several searches undertaken since the original proposal in 1960, although at the time of this writing none have been confirmed (see, e.g., Bradbury 2001a; Tilgner and Heinrichsen 1998; Timofeev, Kardashev and Promyslov 2000). The idea of a 'Dyson shell', or 'Dyson sphere', was subsequently taken up and expanded upon by the astrophysicist Nikolai Kardashev (1964), who conceived of technological civilizations as being characterizable on a three-level scale with regard to their ability to use and control energy.

**The Kardashev Scale**

Kardashev's initial schema has frequently been revised and refined by many others in the five decades since it was first proposed. In brief, it is (e.g., Sagan 1973: 233-234):

- **Type I: planetary.** A Type I civilisation is one which makes use of use all of the available energy of its planet, estimated to be on the order of $10^{16}$ watts (i.e. 10,000,000,000,000,000 W), or $10 \times 10^{15}$ W = 10 PW (petawatts). This would include harnessing, for example, tidal, thermal, atmospheric, nuclear, fossil, internal geothermal and other planetary sources of energy.
- **Type II: stellar.** A Type II civilisation is one which harnesses all of the energy output of its star, something on the order of $10^{26}$ W = $100 \times 10^{24}$ W = 100 YW (yottawatts). This includes collecting all of the radiant energy of the star, and might perhaps even include harnessing the energy contained in its gravitational field.
- **Type III: galactic.** A Type III civilisation is one which has managed to harness the energy of an entire galaxy, something like $10^{36}$ W, although because galaxies vary considerably in size, this figure is somewhat variable. A civilisation capable of using energy at this scale could probably make itself visible, if it chose to, throughout most of the observable universe.

The energy difference between adjacent types is ten orders of magnitude—a factor of 10 billion (i.e. $10^{10}$). Astronomer Carl Sagan suggested (1973: 234) that a decimal interpolation be introduced between the main levels, whereby each factor of 0.1 represents a ten-fold increase on the previous level. Thus, a Type I.5 civilisation uses 10 times more energy than a Type I.4, which uses 10 times more than a Type I.3 and so on. In this view, Earth is usually considered to be an approximately Type 0.7 civilisation.

It is fairly simple to state the characterization of a civilization as 'using energy on a galactic scale', the definition of Type III, but it is not quite so simple to imagine what that situation might entail in terms of artificial structures we might be able to detect.

Does it imply a vast system of beacons, each pulsing out transmissions in the tens of millions of yottawatts range, or would it be something more subtle, such as the ability to move whole star systems around at will, in order to re-configure the wider structure of the galaxy? Is the energy usage expended in a single or small number of artefacts or activities, or is it spread out over innumerable activities whose aggregate total is of the order of magnitude considered galactic in scale?

In the Kardashev scheme, Dyson's idea of re-engineering a star or star system—now usually called 'astroengineering'—is considered to be an example of a Type II civilization. The nature and possible structure of Type III civilizations has received somewhat less attention, although there have been some researchers who have thought along these lines (e.g., Annis 1999; Bradbury et al. 2011; Carrigan 2012; Ćirković 2006). When this question has been considered in the literature at all—which does not appear to have been often—it has usually looked to expanding the scale of Type II civilizations into the galactic context. For example, Carrigan (2012) wrote of 'Fermi bubbles' or 'voids' as places where there is an apparent dearth of visible stars due to the existence of large numbers of Type II civilizations or Dyson Spheres. Such a void or bubble of reduced optical stellar density could, in principle, be detectable by our instruments, owing to the infra-red blackbody radiation signature it would still emit, combined with the unusual appearance that such a structure would produce. However, Carrigan also noted that detection of such voids in spiral galaxies is somewhat more difficult than in elliptical galaxies, owing to the presence of comparable voids that are naturally found in spiral galaxies. The few efforts made to date do not appear to have found definitive examples of galaxy-scale artificial activity (e.g., Annis 1999), but it would be very interesting to search the literature more exhaustively than has been possible for this paper.

**What would a Type III civilization actually look like?**

The ideas to be presented here arose in part from asking the question 'I wonder what a Type III civilization would actually look like?' as well as from some related exploratory investigations into the parameter space for possible scenarios of 'contact'—the usual shorthand term for the discovery of extra-terrestrial life, including intelligence. That paper is currently in preparation (where the technique is described in more detail), but, in brief, the 'scenario space' of contact—that is, the range of possible scenarios under which contact might occur—is assumed to be characterizable by several parameters, including (among others): the nature of the entity (biological, post-biological, hybrid); the complexity of the entity (simple, complex, intelligent); the form of 'signal' (electromagnetic, artefactual); the intentionality of the signal (deliberate, incidental); and the Kardashev type (0, I, II, III). One can see that the initial form of orthodox SETI (i.e., deliberate or incidental leakage radio signals), later forms of orthodox SETI (e.g., deliberate optical signals), and Dysonian SETI (incidental artefact-producing activities) are all accommodated in the parameters characterizing the form and intentionality of the 'signal'. The meaning of 'artefactual' is deliberately left somewhat open so as to encompass Dyson's own suggestion in his Rule 1 of looking for 'artificial activities', and is taken to mean any objects or artefacts produced by any such 'artificial activities'.

This parameterization can be expanded into a many-dimensional (one dimension for each parameter) combinatorial 'morphological space', following a technique devised by Fritz Zwicky in the early part of last century, whereby every parameter value is systematically and exhaustively combined with every other parameter value for all parameters (Zwicky 1967, 1969). This results in a very large number of possible 'configurations', numerically equal to the product of the number of parameter values of all parameters. Zwicky used this technique to great effect in his scientific work (e.g., Zwicky 1947, 1948). Every distinct configuration of parameter values could, in principle, be examined for its characteristics, although in practice not all configurations necessarily appear as 'solutions' because some parameter value pairs may not be mutually 'consistent' and would thereby be excluded from the total 'solution space' (see Voros 2009 for a more detailed explanation of this method). By way of illustration, our own cultural leakage signals—e.g., *I Love Lucy*—can be characterized by the following set of the above parameter values: biological, intelligent (or so *we* might think!), electromagnetic, incidental (i.e., unintentional signalling), Type ~0. More colloquially, this might be rendered as: incidental electromagnetic signals (i.e., 'leakage') from an intelligent, biological, Type ~0 civilization.

For our purposes here, it suffices to say that by combining different classes of parameter value it is possible to systematically generate ideas for different potential scenarios for further consideration and investigation. One of these configuration classes (by which is meant that some parameters are left 'free' without being assigned a specific definite value so as to describe a range of related configurations) was as follows: intelligent, artefactual, incidental, Type III, with the nature of the entity left open. This may be characterized more colloquially as: a galaxy-scale artefact created by, or perhaps galaxy-scale artificial activities undertaken by, some form of intelligent entity going about its own business; in other words, galaxy-scale macro-engineering.

In what follows, I would like to consider two possibilities for how galactic-scale changes brought about by macro-engineering activities might manifest in terms of structures we might be able to detect over intergalactic distances. The timeframe for this scale of engineering is probably rather long, and might run to many tens of millions, or perhaps even hundreds of millions of years. Dick (2003) has called thinking on these immense timescales 'Stapledonian' and suggests that such long-term thinking is a necessity when considering the question of intelligence in the universe. In the spirit of Dyson's rules, however, we are only concerned here with artefacts or artificial activities that we could actually detect over intergalactic distances and not with what might be considered an 'average' level of macro-engineering for a 'typical' engineering species. That is, we are concerned only with the biggest, most astonishingly vast engineering projects of which we can possibly conceive—so, we will be thinking along the lines of what Ćirković (2006) has characterized as 'macro-engineering in the galactic context'.

**Multi-system exponentiating Dysonian astroengineering**

Firstly, let us imagine a long-lived Dyson/Type II species branching out from its initial home system, which would likely have been somewhere in the Galactic Habitable Zone (GHZ) (see, e.g., Prantzos 2008) where they most likely first arose as a biological species—although by this stage they may well have moved to a 'post-biological' form (Dick 2003, 2009). Over numerous iterations, new stellar systems are reached by a vanguard group sent from an existing 'Dyson-ified' system and subsequently engineered into new Type II systems, from which new groups are sent out, and so on. This is clearly a geometrically exponentiating process so that, over time, there will end up being a *very* large number of Type II/Dyson civilizations, spreading out in a roughly spherical bubble from the home system. This is similar to the scenario of Fermi voids or bubbles that Carrigan (2012) has imagined. However, we can push this idea a bit further by considering what an initially-spiral galaxy might look like some way further along the exponentiating process. Owing to slight differences in the orbital speed of star systems around the galactic centre due to differences in radial distance from the centre, this expanding bubble is not likely to remain completely spherical, and may end up getting progressively 'smeared out' by differential rotation at the different radii (an admittedly fairly small effect). Over Stapledonian time-frames, however, comparable to several galactic rotations (say $\sim 10^9$ years), this process, which might otherwise have led to a (so to speak) Fermi 'arc' around the centre of the galaxy, could very likely end up filling out into a fully-blown 'gap' that completely separates the galactic core from the rest of the outer stellar disk, through the still-continuing process of diffusion from existing engineered systems into new un-engineered ones.

The radii of this 'gap' in the galactic disk would likely be determined, respectively, by the intensity of the radiative flux from the galactic core or bulge, for the inner, and by the availability of metal-rich stars which contain planetary and other materials suitable for disassembly and re-use for engineering purposes, for the outer. The resulting gap may end up encompassing much of what might have been considered the GHZ of the galaxy, at least to the outer radius in the galactic disk. It is also possible that the inner radius might extend even further inward towards the centre of the galaxy, if the species 'goes post-biological' and no longer has to worry about the effects of what would otherwise be lethal environmental conditions for a biological species. In this case, 'post-biological' is a general term which may be inferred to include machine-based intelligence, or an intelligence based on a technological/artefactual substrate. In SETI, this situation is sometimes referred to colloquially as the question of these intelligences either *having* machines or *being* machines, and the question itself is sometimes regarded as mere hair-splitting.

**Galactic-structural macro-engineering and stellar-system removal**

In the second possibility, we imagine a long-lived most-likely post-biological species inhabiting a spiral galaxy that either does away altogether with its earlier exponentiating Dysonian astroengineering program in favour of, or perhaps moves directly to, the even grander project of seeking to re-engineer the spiral-galactic structure as a whole. If this species transitions to a post-biological form while still planet-bound, or relatively early into a nascent exponentiating astroengineering phase, then this latter trajectory may perhaps be more probable.

Analogously with Dyson's original astroengineering proposal to harvest stellar energy, although on a much larger scale, this species decides that it wants to directly access and capture all of the luminous energetic flux emanating from the entire galactic core or bulge. Unfortunately, there is usually a considerable amount of intervening material in a typical spiral galaxy which occludes some of this radiant energy from regions further out in the galactic disk by the absorption of some wavelengths—this is why we ourselves do not see the centre of the Milky Way from Earth in visible light. A post-biological species would likely not have need of planets as habitat, and would most probably be able to exist in interstellar space, absorbing the radiant energy directly in a way analogous to our current solar panels, although undoubtedly much more efficiently. Such a species would not be constrained by biological timelines, and—if our own considerations about the implications of The Singularity on Earth are anything to go by (e.g., Eden et al. 2012; Kurzweil 2006; Smart 2003; Tucker 2006; Vinge 1993)—would effectively become immortal, subject only to accidental destruction, or a conscious decision to power-down.

An effectively-immortal species which wants to gain access to as much of the galactic centre's radiant energy as possible without it being degraded due to absorption and re-emission would likely consider clearing out the intervening material between itself and the galactic core/bulge. This would be macro-engineering on a truly galactic scale. One can imagine a number of possible scenarios for how this might proceed. The asymptotic end-state of these activities on Stapledonian timescales would likely be: a central core of stars, surrounded by a 'gap' in which there are relatively much fewer or perhaps even no stars or other natural material—and which contained uncounted octillions of post-biological entities orbiting the galactic core absorbing the unimpeded radiant flux as they go about their unfathomably post-biological business—with a ring of stars further out remaining from the initial structure of the spiral galaxy. The removal of intervening materials might include combinations of 'lifting' stars entirely for later re-use of their materials (e.g., Criswell 1985), or perhaps simply moving entire star systems further out into the stellar ring region, by means of some form of propulsion such as Shkadov thrusters or related 'stellar engines' (e.g., Badescu and Cathcart 2000, 2006a, 2006b). Or, it might perhaps be through a combination of lifting stars into Jovian planet- or brown dwarf-sized non-fusioning agglomerations for more convenient storage and then moving these with gravitationally-bound 'solar sails' utilizing the radiation pressure from the galactic centre. In this case, as well, one ends up with a core of stars surrounded by a quasi-toroidal region devoid of stars, ultimately out to a more distant ring of stars, whose inner radius is determined due to the radiant flux from the core being too weak for the post-biological species to utilize, whether directly or for propulsion. This argument is in direct contrast to the 'migration hypothesis' of Ćirković and Bradbury (2006) who have argued for a mass migration of post-biological species to the *outer* regions of a galaxy, for computing-thermodynamic reasons.

**Galactic structure arising from these macro-engineering projects**

In both of these cases, the galaxy eventually ends up having a core + 'gap' + ring morphology, reminiscent of the planet Saturn, where the apparently-empty 'gap' might merely be comparatively darker due to the presence of a vast number of Dyson spheres, or may actually have been emptied due to the original material having been cleared away—through star lifting, star system re-positioning or similar forms of removal—to make for more open 'living' space for the post-biological entities. We will consider further below some possible empirical observations that could be made of any such candidate galaxy. But for now, of course, the question arises: are there any examples of galaxies that have this morphology? And the answer is: yes, there are.

**Hoag's Object – a lovely ringed galaxy**

The nature and structural characteristics of the beautiful ringed galaxy known as Hoag's Object, which has the formal designation PGC54559, have long been the subject of debate and speculation (e.g., Brosch 1987; Gribbin 1974; Lucas 2002). It was discovered by Arthur Hoag in 1950, who reported it in the scientific literature as a 'peculiar object' (Hoag 1950), hence its common name 'Hoag's Object'. It lies about 600 million light-years away in the constellation Serpens and is something like 100-120,000 light-years across, making it roughly comparable to or slightly bigger than the Milky Way (Brosch 1985; O'Connell, Scargle and Sargent 1974; Schweizer et al. 1987). Detailed analysis shows that the galactic plane is almost directly face-on to us, deviating from perfect alignment by only about 20 degrees or so (Schweizer et al. 1987). The interested reader can see a high-quality Hubble Space Telescope image of Hoag's Object at the Astronomy Picture of the Day web site for August 22, 2010 (Lucas and NASA Hubble Heritage Team 2010). There are several other such 'Hoag-type' galaxies known (O'Connell et al. 1974; Wakamatsu 1990), including one that is, coincidentally, visible through the gap feature in Hoag's Object itself (Lucas and NASA Hubble Heritage Team 2010).

Hoag initially thought it might be a possible example of gravitational lensing, the ring being an optical effect caused by the bending of light from a more-distant galaxy by an intervening elliptical galaxy located by chance directly in line-of-sight between us and the more distant one. Later spectroscopic work showed this not to be the case (O'Connell et al. 1974), and as both the core and the ring appear to have the same redshift, they are almost certainly co-located (Schweizer et al. 1987). A variety of other hypotheses have also been proposed for the origin of this lovely galaxy. They include: a 'bulls-eye' type collision between two passing galaxies—however there does not appear to be any sign of the putative 'bullet' galaxy in the vicinity (Schweizer et al. 1987); a dynamical instability in what was previously a barred-spiral galaxy, which case can be recovered from adjusting the parameters modelling the galactic dynamics in certain ways (Brosch 1985; Freeman, Howard and Byrd 2010); an accretion event, wherein the object we see is a late stage in the coalescing process of two colliding galaxies merging into one system (Schweizer et al. 1987); and, more recently, that the structure we see can be modelled by a particular type of pressure wave in a self-gravitating gas (Pronko 2006). The last three of these appear to remain viable hypotheses.

However, given our use here of Dysonian thinking over Stapledonian timeframes, and the sometimes finely-tuned adjustments in the parameters that appear to be necessary to recover the structure of Hoag's Object via models of natural processes, what if we instead ask the question that is now almost begging to be asked: Is Hoag's Object actually an example of galaxy-scale macro-engineering? Or, put more simply:

**Is Hoag's Object an artefact?**

Having asked this question, of course, the next step is to consider how we might go about answering it. This requires thinking about potential empirical observations that could be undertaken in order to look for 'signatures' that would indicate artificial activities rather than be explicable as due solely to natural processes.

There appear to be at least four empirical observations that could be made with respect to the question of the artificiality or otherwise of Hoag's Object. They range from rather less direct to considerably more direct, and are as follows:

1. **The distances from the centre of the galaxy of the major structural discontinuities**—the outer core/inner gap, and outer gap/inner ring radii—**with regard to what these distances might be expected to be from theoretical considerations arising from different scenarios leading to the core/gap/ring morphology**. If stars or stars systems *are* being moved outward by radiation pressure, for example, then there will be a certain radial distance at which the inward gravitational attraction—set by the core mass—and the outward radiation pressure—set by the core luminosity—are in balance. This would form the boundary of the core and gap. Similarly, the outer gap radius could also be limited by the intensity of radiation pressure for moving material outwards. However, the outer radius of the gap might not be so strongly constrained if the stars are being moved using Shkadov thrusters. If the 'empty gap' appearance is caused by Dyson spheres rather than lack of material, there may perhaps be a different implied radius for the core/gap boundary. Comparisons of different theoretical values, obtained from imagining different scenarios, with the empirical values obtained from direct observation might reveal some interesting correspondences. This may well require more accurate observational data than currently appear to exist.

2. **The spectral profile of the 'gap'**. The difference between a 'gap' consisting of, say, almost-empty space containing octillions of post-biological entities as compared to a volume of space filled with billions of Dyson Spheres should, in principle, be discernible, but in reality might be very difficult to determine conclusively. If there are Dyson spheres of the type considered by Dyson himself, these will emit blackbody radiation consistent with a temperature of ~300 K. However, if the engineering is very advanced—and we almost must assume it to be—then the efficiencies possible by use of a structure similar to, say, a 'Matrioshka Brain' (Bradbury 2001b) could produce waste heat with a blackbody radiation profile close to the temperature of the cosmic background radiation of space itself, ~3 K. Similar considerations for thermodynamic efficiency would probably also drive the construction of the substrate for the post-biological entities. This could make it very difficult to distinguish such structures from background empty space. In this instance, possible occlusion or diminution of radiation intensity from beyond the galaxy due to intervening absorptive material might be one way to probe the nature of the 'gap'.

3. **The metallicity profile and chemical composition of the ring**, with respect to anything 'unusual' compared to what is expected of the 'typical' composition of the interstellar medium (ISM) in a galaxy in which the normal processes of stellar evolution are occurring. A species that is converting gap-region star systems into Dyson spheres and not moving materials further outward would not therefore alter the composition of the ring region to any appreciable degree beyond what would be expected from the normal processes of enrichment over time of the ISM in a spiral galaxy. However, if the species *is* moving gap-region materials further outward, then this probably *would* alter the chemical composition and metallicity profile of the ring region and such an 'anomalous' composition might be detectable via spectroscopic observation. The ring structure in Hoag's Object also shows what Schweizer and co-workers (1987) characterized as an 'osculating braid', a smaller brighter ring within the main ring, touching the inner and outer edges of the main ring at different places; it is clearly visible in the Hubble image mentioned above. The precise nature of this 'braid' is of some interest. It has the appearance of what are in other galaxies considered to be regions of relatively new star formation, although, if so, how so much star formation has managed to be so apparently spatially synchronized raises an intriguing speculation. Is the braid simply a pressure-shock effect caused by multiple and perhaps cascading supernovae events, or could it be due to the deliberate synchronized 'seeding' of new star systems? And to what end? It is possible to imagine that an effectively-immortal post-biological species which is already moving material into the outer ring region might decide as part of this project to undertake a further program of seeding the creation of new stars with the materials so displaced in order to create the potential for the emergence of new biological species. In other words, to perhaps 'cultivate' the sort of conditions conducive for the arising of new biological species (in a region of the galaxy for which they themselves have no direct use) perhaps for subsequent longitudinal observation and study. Or it might simply be to produce an aesthetic effect that changes relative position within the ring over very long timescales, possibly even as a signal to other galaxies. The prospect of effective immortality might require commensurately long-term projects to keep one occupied over the aeons.

4. **The existence of a time-keeping signal beacon** at or near the galactic core. If there are large-scale engineering activities going on which could be up to many tens of thousands of light-years apart and which may require some type of co-ordination, then it would be useful to have a time signal that would act as the standard clock by which these activities could be synchronized—a kind of Galactic Mean Time, as it were (Shostak 1999). A logical place for such a beacon would be at the galactic core, or perhaps immediately nearby, slightly above the plane of the galaxy, as the exact centre may not be feasible due to a black hole or other sources of possible interference with what would need to be a reliable signal. It is most unlikely that such a beacon would be broadcast isotropically—that would be a distinctly inefficient use of energy. An immortal species that is re-engineering a galaxy over Stapledonian timescales is quite likely to be somewhat careful in its use of energy, perhaps even frugal, as it would probably be thinking very much of the long term—and it would certainly have the luxury of time enough to use the most frugally efficient means possible. As such, the signal would most likely be directed mostly along the galactic plane (e.g., Shostak 2011: 363) to the regions where it would be needed. Given that Hoag's Object is almost directly face-on to us, this makes the detection of any potential beacon signal of this kind somewhat difficult, although one might hope for some re-emission scattering echoes being deflected in our direction. However, if there were other activities being undertaken further out in the galactic halo, the signal might then be broadcast somewhat more widely, so we might possibly get lucky. Needless to say, a very sensitive receiver would be required for carrying out such an observing program.

**Concluding remarks**

This paper has been an attempt to apply 'Dysonian' thinking to the question of what galaxy-scale macro-engineering might look like when undertaken over 'Stapledonian' cosmological timeframes by intelligent species that are long-lived enough to do so, and which have very probably transitioned to an effectively-immortal 'post-biological' form (Bradbury et al. 2011; Ćirković 2006; Dick 2003).

By generalizing Dyson's original idea—an engineering species interested in harnessing ever-more amounts of radiant energy—from a single solar system to an entire galaxy, we arrived at the intriguing notion of a purposely re-engineered galaxy eventually having a core + 'gap' + ring morphology, somewhat reminiscent of the planet Saturn. When we look to the sky, we find that there are indeed several examples of such galaxies, the most well-known 'type specimen' of which is Hoag's Object, PGC54559. The unusual structure of this beautiful galaxy has long been remarked upon, and the attempt to resolve the question of its origin has seen a variety of hypotheses advanced based on the assumption of natural processes. Here a different question was posed concerning its origin—namely, whether it might actually be the result of artificial activities. Thus we asked: Is Hoag's Object an example of galaxy-scale macro-engineering? Several theoretical considerations were discussed and four specific empirical observations were suggested that could be carried out in order to begin to investigate this wonderfully beguiling research question.

Mounting a search for evidence of galaxy-scale macro-engineering, and the thinking required to seriously contemplate the possible forms such projects might take in order to be able to do so, could be one way to help us think much longer-term—something humankind would seem to be in desperate need of right now. If nothing else, simply entertaining the idea that someone somewhere in the Universe might have been able to successfully navigate the dangerous time Carl Sagan called 'technological adolescence' can give us some hope in our own ability to do the same at this critical point in the history of our species. Better yet, finding a definite example of such a success could be the very stimulus we need that prompts us to begin to take our future seriously enough to guide it consciously, responsibly and foresightfully. A search for evidence of this kind would be relatively inexpensive to conduct. But it just might end up being an immeasurably valuable—perhaps even absolutely priceless—piece of information to possess. It couldn't hurt to have a careful look... .

*